\documentclass[showpacs,preprintnumbers,amssymb]{revtex4}
\usepackage{graphicx}
\usepackage{epsfig}
\usepackage{dcolumn}
\usepackage{bm}

\def\be{\begin{equation}}
\def\ee{\end{equation}}
\def\bea{\begin{eqnarray}}
\def\eea{\end{eqnarray}}
\def\beqa{\begin{eqnarray*}}
\def\eeqa{\end{eqnarray*}}
\def\nnb{\nonumber}

\def\gsim{\lower0.5ex\hbox{$\:\buildrel >\over\sim\:$}}
\def\lsim{\lower0.5ex\hbox{$\:\buildrel <\over\sim\:$}}

\def \n{\noindent}

\begin{document}
\title{Lepton-flavor-violating decays $L\to l\gamma\gamma$\\ 
       as a new probe of supersymmetry with broken R-parity}
\thanks{Work supported in part by the EC 5th framework, 
	contract number HPMF-CT-2002-01663}
\author{Alexander Gemintern}
\email[phcga@physics.technion.ac.il]{}
\affiliation{Technion--Israel Institute of Technology,\\
             32000 Haifa, Israel}
\author{Shaouly Bar-Shalom}
\email[shaouly@physics.technion.ac.il]{}
\affiliation{Technion--Israel Institute of Technology,\\
             32000 Haifa, Israel}
\author{Gad Eilam}
\email[eilam@physics.technion.ac.il]{}
\affiliation{Technion--Israel Institute of Technology,\\
             32000 Haifa, Israel}
\author{Frank Krauss}
\email[Frank.Krauss@cern.ch]{}
\affiliation{Theory Division, CERN\\
             CH-1211 Geneva 23, Switzerland}
\date{\today}
\preprint{CERN-TH/2003-034}

\begin{abstract}
Lepton-flavor-violating decays of the type 
$L\to l\gamma\gamma$ ($\mu\to e\gamma\gamma$, 
$\tau\to e\gamma\gamma$, and $\tau\to \mu\gamma\gamma$) 
are proposed as new probes of R-parity-violating 
supersymmetry. Non-penguin diagrams with a sneutrino that 
decays into two photons via a triangle graph might 
trigger such decays even in the absence of the corresponding 
radiative decays into one photon only, e.g. $\mu\to e\gamma$. 
Thus, processes of the type $L\to l\gamma\gamma$ may provide 
an independent probe of new flavor physics.  
\end{abstract}


\maketitle

Supersymmetry (SUSY) is one of the most promising candidates 
for new physics, curing some of the shortcomings of the 
Standard Model (SM), for instance, the emergence of quadratic 
divergences in the Higgs sector \cite{SUSY}. Despite its 
appealing theoretical features, however, there is so far no 
experimental evidence for SUSY up to the electroweak scale. 
Apart from direct evidence, i.e. direct production and decays 
of SUSY particles, indirect probes could be employed to search 
for SUSY, i.e. processes in which the SUSY particles emerge 
virtually either in loops or as tree-level mediators. In this 
respect Flavor-Changing Neutral Current (FCNC) or 
Lepton-Flavor-Violating (LFV) processes are a natural 
searching ground for SUSY, since the SM background to such 
processes is either loop-suppressed (FCNC) or does not 
exist (LFV). 

In this paper the effects of new physics on a class of LFV 
processes, $L\to l\gamma\gamma$, are investigated, in which a 
lepton decays radiatively into a lighter one plus two photons. 
As an example, R-Parity-Violating (RPV) SUSY is discussed. 

In the conventional RPV framework, the SUSY superpotential is 
supplemented with new RPV interaction terms that are either 
baryon- or lepton-number-violating \cite{Drei, RPV}. For the 
purpose of this paper, only the following lepton-number-\-violating 
operators are relevant:

\begin{eqnarray}\label{LRPV}
{\cal W}_{RPV} \supset 
\frac{1}{2} \lambda_{ijk} \epsilon_{ab} 
            {\hat L}^a_i {\hat L}^b_j {\hat E}_k^c +
\lambda_{ijk}^\prime \epsilon_{ab} 
            {\hat L}^a_i {\hat Q}^b_j {\hat D}_k^c ~,
\end{eqnarray} 

\n where ${\hat Q}$ and ${\hat L}$ are SU(2) doublet quark and 
lepton supermultiplets, respectively, and ${\hat D}^c$ and 
${\hat E}^c$ denote the SU(2) singlet down-type quark and lepton 
supermultiplet; $i,j,k=1,2,3$ label the generations and $a,b$ 
are the SU(2) indices that are responsible for the antisymmetric 
property $\lambda_{ijk}=-\lambda_{jik}$.

\begin{figure}
\includegraphics[width=5cm,angle=270]{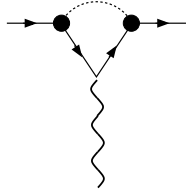}
\caption{\label{peng1} 
         A penguin diagram responsible for the decays
         $L\to l\gamma$ in RPV SUSY. The photon can couple to
         any line provided it is charged. RPV couplings
         are symbolized by a blob.}
\end{figure}

Within the framework of RPV SUSY, $\mu\to e \gamma$ 
(and $\tau\to e\gamma$, $\tau\to\mu\gamma$) can be mediated by 
penguin-like diagrams of the type depicted in Fig. \ref{peng1}. 
Depending on the RPV couplings involved, the particles running in 
the loop are either quark--squark, or lepton--slepton. The process 
$L\to l\gamma$, where $L$ and $l$ are leptons of generations 
$i$ and $j$, respectively, is then proportional to 
either $\lambda_{imn}\lambda_{jmn}$ or $\lambda_{mni}\lambda_{mnj}$ 
or $\lambda'_{imn}\lambda'_{jmn}$
\cite{Chai}. 
 
\begin{figure}
\includegraphics[width=7cm]{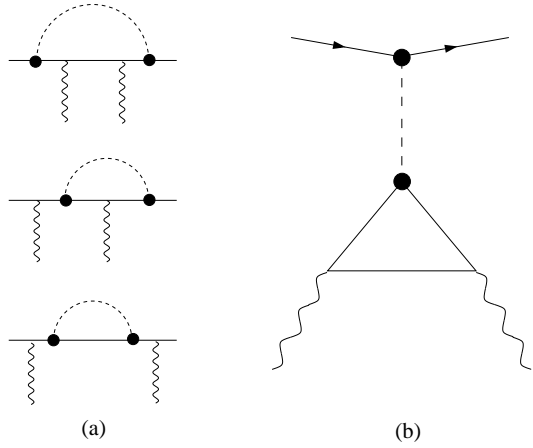}
\caption{\label{peng2} 
         The topology of (a) $\lambda$-{\it reducible} and
         (b) $\lambda$-{\it irreducible} diagrams for 
         $L\to l\gamma\gamma$.}
\end{figure}

So far, very little attention has been devoted to 
$L\to l\gamma\gamma$, under the assumption that it is always 
suppressed with respect to $L\to l\gamma$. 
An exception to this rule is discussed in \cite{leptoq}, where a leptoquark 
model in which $L\to l\gamma\gamma$ is much larger than $L\to l\gamma$ is
presented. Note that their model yields $BR(\mu\to e\gamma\gamma)<10^{-18}$;
see also \cite{Bowman}. In general, an additional photon can easily be 
accommodated - one just has to add a photon 
to the diagram in Fig.~\ref{peng1} and obtain the {\it $\lambda$-reducible} 
diagrams shown in Fig. \ref{peng2}a (they are reducible in the sense 
that, after removing a photon line from $L \to l \gamma \gamma$, a
legitimate RPV diagram for $L \to l \gamma$ is obtained). Obviously 
processes mediated through such diagrams do not provide any further 
information on the corresponding RPV couplings, since they differ 
only by kinematic factors and by an extra electromagnetic coupling.

However, it is quite possible that the decay $L\to l\gamma$ does 
not exist, or its rate is too small to be observed, whereas the decay 
$L\to l\gamma\gamma$ still has an appreciable rate. This could be the 
case if, for instance, all products of the type $\lambda\lambda$ and 
$\lambda'\lambda'$ are much smaller than products of the type 
$\lambda\lambda'$, in which case the {\it $\lambda$-irreducible} 
topology of Fig.~\ref{peng2}b may give rise to a much larger rate 
than the {\it $\lambda$-reducible} diagrams of Fig.~\ref{peng2}a.

For example, if $\lambda_{122}$ and $\lambda'_{233}$ are the only
non-vanishing lambdas, then $\Gamma(\mu\to e\gamma) = 0$ while 
$\Gamma(\mu\to e\gamma\gamma)\ne 0$. Accordingly, such RPV coupling 
product combinations can be detected or constrained by the measured 
upper limit on the branching ratio of $\mu\to e\gamma\gamma$, which is 
at present $7.2\times 10^{-11}$ \cite{PDG,experiment2}. No experimental
limits exist so far for the analogous decays of the $\tau$-lepton. 

There are
proposals to measure (or constrain) the branching 
ratios of such SM-forbidden decays of the $\mu$ down to branching ratios 
of the order of $10^{-14}$ \cite{experiment4}, $10^{-15}$
\cite{experiment5} or even down to $10^{-18}$ \cite{experiment6}.
This highlights the phenomenological impact of 
LFV muon decays.

A general basis of effective operators describing the decays 
$L\to l\gamma\gamma$ with the topology of Fig. \ref{peng2}, 
i.e. $L \to l S_k \to l\gamma \gamma$, will be constructed first.
Here $S_k$ is a scalar particle of type $k$ and $i,j$ are generation indices
of $L,l$, respectively. 
The momentum notation $L(p)\to l(p') \gamma(k)\gamma(k')$ will be used. 

The amplitude can be defined as

\be\label{OPE}
{\cal M}_{L \rightarrow l \gamma\gamma}=
\sum_{n=1}^2 \sum_{{\cal P}=L,R} 
             C_{n(ijkm)}^{{\cal P}} {\cal O}_n^{{\cal P}},
\ee
\noindent 
where $m$ is the family index of the 
fermion that runs in the loop, and the index $k$ corresponds to the 
intermediate scalar. Also, $C_{n(ijkm)}^L,~C_{n(ijkm)}^R$ are 
Wilson-like coefficients with $L(R) = (1-(+)\gamma_5)/2$, and

\bea\label{Operators}
{\cal O}_1^{L} &=& 
\left[\bar{u}_l(p') L u_L(p)\right]\,
      \left(k \cdot k'g^{\lambda\nu}-k'^{\lambda}k^{\nu}\right)
      \epsilon^*_{\lambda}\epsilon^{\prime *}_{\nu}\nnb\\
{\cal O}_2^{L} &=& 
\left[\bar{u}_l(p') L u_L(p)\right]\,
      i\varepsilon^{\lambda\nu\rho\sigma}k_{\rho}k'_{\sigma}
      \epsilon^*_{\lambda}\epsilon^{\prime *}_{\nu}
\eea

are effective operators. If the incoming lepton is right-handed, 
then $L \to R$ in (\ref{Operators}).

In the framework of RPV SUSY with a sneutrino ($\tilde\nu_k$) 
exchange, the Wilson coefficients are given by:   

\begin{widetext}
\bea
C_{1;2(ijkm)}^L;~C_{1;2(ijkm)}^R(\ell_m {\rm ~in~the~loop}) &=&   
   0;~\frac{\alpha}{4\pi}
\frac{i\lambda_{kji}^\star \lambda_{kmm}}{m_{\ell_m} M_{\tilde\nu_k}^2}
   \cdot \left[ f_{1/2}(x);g_{1/2}(x) \right]~, \nnb\\
C_{1;2(ijkm)}^L;~C_{1;2(ijkm)}^R(d_m {\rm ~in~the~loop}) &=& 
   0;~N_cQ_d^2
   \frac{\alpha}{4\pi}
\frac{i\lambda_{kji}^\star \lambda'_{kmm}}{m_{d_m} M_{\tilde\nu_k}^2}
   \cdot \left[ f_{1/2}(x);g_{1/2}(x) \right]~, \label{CLCR}
\eea
\end{widetext}

\noindent where $M_{\tilde\nu_k}$ is the mass of the $k$th 
generation sneutrino and $m_{f}$, $f=\ell~{\rm or}~d$, is the mass
of the fermion of generation $m$ that runs in the loop. Furthermore 
$N_c=3$, $Q_d=1/3$ and the abbreviation $x=2m_{f}^2/k \cdot k'$ is being 
used. For diagrams with a charge-conjugate sneutrino ($\tilde\nu^\star$) 
exchange, in (\ref{CLCR}) $C_{1;2(ijkm)}^L \leftrightarrow C_{1;2(jikm)}^R$
(i.e. with $\lambda_{kji}^\star \to \lambda_{kij}^\star$) and 
$M_{\tilde\nu_k} \to M_{\tilde\nu_k^\star}$.
The functions $f_{1/2}$ and $g_{1/2}$ (originating from the loop 
integral of the triangle) are given by

\bea\label{Functions}
f_{1/2}(x) &=& -2x\left[1+(1-x)
               \arcsin^2\left(\frac{1}{\sqrt{x}}\right)\right]\,, 
               \nnb\\
g_{1/2}(x) &=& 2x\arcsin^2\left(\frac{1}{\sqrt{x}}\right) ~.
\eea

The emergence of these functions reflects the different structure of 
the integral related to the parity of the couplings of the fermion in 
the loop. The subscripts refer to the spin of the particle running in the 
triangle. Note that when $x<1$, the 
$\arcsin\left(1/\sqrt{x}\right)$ develops an imaginary part, 
which corresponds to the loop particle residing on its mass shell. 

In what follows, results will be given for 
both constituent and current light quark masses:
$m_d=300$ MeV, $m_s=450$ MeV, $m_b=4.5$ GeV, and 
$m_d=10$ MeV, $m_s=120$ MeV, $m_b=4.2$ GeV, respectively. 
We believe that using constituent quark masses makes more sense,
since for our case here, it would allow the decay $\mu \to e d \bar{d}$, for
current $d$ quarks \cite{Eeg}.
 
For the case of the muon decay, $\mu \to e \gamma \gamma$, and for either
down quarks with constituent masses or for the $\tau$-lepton running in 
the loop, the relation $m_f^2 > m_\mu^2 > k \cdot k'$ holds. Then, the 
limit $m_f^2\gg k\cdot k'$ ($m_f=m_d,~m_s,~m_b$ or $m_\tau$) provides a good 
approximation. Therefore, the loop integral functions can be replaced 
with their $x\to\infty$ limits, $f_{1/2}(x\to\infty)=-\frac{4}{3}$ and 
$g_{1/2}(x\to\infty)=2$. Using this approximation, decay widths are 
given by

\begin{widetext}
\begin{eqnarray}
\Gamma(\mu \to e \gamma \gamma)\mid_{m_f^2 \gg k \cdot k'} \simeq
A_{RPV}^{f S} \left(\frac{\alpha}{\pi}\right)^2\frac{1}{256\pi^3}
\frac{13m_{\mu}^7} {4320 m_{f}^2M_{S}^4},
\end{eqnarray}
\end{widetext}

\noindent where $f=d_i$ ($i=1,2,3$ for $d,s$ or $b$, respectively) or 
$f=\tau$ is the fermion in the loop and $S=\tilde\nu_k$ or 
$\tilde\nu^\star_k$ ($k=1,2$ or 3), is the scalar responsible for the 
flavor changing transition $\mu \to e$. Also
\begin{widetext}
\begin{eqnarray}
A_{RPV}^{\tau \tilde\nu_k}= \mid \lambda_{k12} \lambda_{k33}\mid^2 ~,~
A_{RPV}^{\tau \tilde\nu_k^\star}= \mid \lambda_{k21} \lambda_{k33}\mid^2 
~,~
A_{RPV}^{d_i \tilde\nu_k}= \mid \lambda_{k12} \lambda'_{kii}\mid^2/9 ~,~
A_{RPV}^{d_i \tilde\nu_k^\star}= \mid \lambda_{k21} \lambda'_{kii}\mid^2/9 ~.
\end{eqnarray}
\end{widetext}

The branching ratios scaled by the relevant RPV couplings for single 
particles running in the triangle are given in Table \ref{results}. In the
spirit of this paper, only such down quarks and leptons running in the 
triangle are considered that would not give rise to the decay 
$L\to l \gamma$, i.e., only those leptons that cannot generate the 
$\lambda$-{\it reducible}-type diagrams.

Comparing the current limit $BR(\mu \to e \gamma \gamma)<7.2\times 10^{-11}$ 
\cite{PDG} with the estimate for the maximal possible branching ratio for 
this decay mode in Table \ref{results}, in conjunction with the existing 
bounds on the relevant RPV coupling products as given in Table 
\ref{lambdas}, it can be argued that, at present, $\mu \to e \gamma \gamma$ 
does not impose any new constraints on the RPV SUSY parameter space.
However, the situation will improve with the much more stringent 
experimental constraints that can be anticipated:
on the third row of Table \ref{lambdas} the expected sensitivity of 
$\mu \to e \gamma \gamma$ to the relevant RPV coupling products is exhibited, 
assuming that future experiments will be sensitive to 
$BR(\mu \to e \gamma \gamma) \lsim 10^{-14}$, see e.g. \cite{experiment4}.

For the case of $\tau$-decays, taking the existing bounds on the 
relevant RPV coupling products into account, the largest branching ratio is 
obtained for $\tau \to \mu \gamma \gamma$ with an $s$-quark in the loop. 
In particular, for $|\lambda_{233} \lambda'_{322}|=0.01$ (its upper bound, 
see \cite{SK-02}), $m_s=450$ MeV and $M_{\tilde\nu}=100$ GeV, 
$BR(\tau \to \mu \gamma \gamma) \sim 10^{-10}$. This result is by far smaller 
than the typical present limits on $\tau$ decay branching ratios, which are 
of the order of $10^{-7}$, see e.g. \cite{CLEO}. Moreover, 
planned experiments \cite{Turks}, which will produce about 
$5 \times 10^8$ taus per year, will also be insensitive to the decays 
$\tau\to l\gamma\gamma$.

\begin{table*}[htb]
\begin{tabular}{|c|c|c|c|c|c|c|} \hline
Loop& \multicolumn{2}{c|}{$BR(\mu\rightarrow e\gamma\gamma)/|\lambda\lambda'|^2$} & 
\multicolumn{2}{c|}{$BR(\tau\rightarrow e\gamma\gamma)/|\lambda \lambda'|^2$} & 
\multicolumn{2}{c|}{$BR(\tau\rightarrow \mu\gamma\gamma)/|\lambda \lambda'|^2$} \\ 
\cline{2-7}
particle & $\tilde\nu_k$       & $\tilde\nu^\star_k$ & $\tilde\nu_k$ & 
           $\tilde\nu^\star_k$ &$ \tilde\nu_k$       & $\tilde\nu^\star_k$ \\ 
$\Downarrow$& 
$(\lambda_{k12}^\star \lambda'_{kmm})$ & 
$(\lambda_{k21}^\star \lambda'_{kmm})$ &
$(\lambda_{k13}^\star \lambda'_{kmm})$ & 
$(\lambda_{k31}^\star \lambda'_{kmm})$ &
$(\lambda_{k23}^\star \lambda'_{kmm})$ & 
$(\lambda_{k32}^\star \lambda'_{kmm})$ \\ \hline \hline
$d$ ($m=1$) & \multicolumn{2}{c|}{$1.3\times 10^{-8}$ ($4.1\times
10^{-6}$)} &
\multicolumn{2}{c|}{$8.7\times 10^{-7}$ ($5.6\times 10^{-8}$)}&
\multicolumn{2}{c|}{$8.5\times 10^{-7}$ ($5.5\times 10^{-8}$)}
\\ \hline
$s$ ($m=2$) & \multicolumn{2}{c|}{$5.5\times 10^{-9}$ ($8.2\times
10^{-8}$)} &
\multicolumn{2}{c|}{$7.6\times 10^{-7}$ ($6.0\times 10^{-7}$)}&
\multicolumn{2}{c|}{$7.4\times 10^{-7}$ ($5.9\times 10^{-7}$)}
\\ \hline
$b$ ($m=3$) & \multicolumn{2}{c|}{$5.5\times 10^{-11}$ ($6.3\times
10^{-11}$)} &
\multicolumn{2}{c|}{$2.8\times 10^{-9}$ ($3.2\times 10^{-9}$)}&
\multicolumn{2}{c|}{$2.7\times 10^{-9}$ ($3.1\times 10^{-9}$)}
\\ \hline \multicolumn{7}{c}{$$}\\
\end{tabular}
\begin{tabular}{|c|c|c|c|c|c|c|} \hline
Loop& \multicolumn{2}{c|}{$BR(\mu\rightarrow e\gamma\gamma)/|\lambda\lambda|^2$} & 
\multicolumn{2}{c|}{$BR(\tau\rightarrow e\gamma\gamma)/|\lambda \lambda|^2$} & 
\multicolumn{2}{c|}{$BR(\tau\rightarrow \mu\gamma\gamma)/|\lambda \lambda|^2$} \\ 
\cline{2-7}
particle & $\tilde\nu_k$       & $\tilde\nu^\star_k$ & $\tilde\nu_k$ 
         & $\tilde\nu^\star_k$ &$ \tilde\nu_k$       & $\tilde\nu^\star_k$ \\ 
$\Downarrow$& 
$(\lambda_{k12}^\star \lambda_{kmm})$ & 
$(\lambda_{k21}^\star \lambda_{kmm})$ &
$(\lambda_{k13}^\star \lambda_{kmm})$ & 
$(\lambda_{k31}^\star \lambda_{kmm})$ &
$(\lambda_{k33}^\star \lambda_{kmm})$ & 
$(\lambda_{k32}^\star \lambda_{kmm})$ \\ \hline \hline
$e$ ($m=1$) & \multicolumn{2}{c|}{$\mu\to e\gamma$ possible} &

\multicolumn{2}{c|}{$\tau\to e\gamma$ possible} & 
\multicolumn{2}{c|}{$8.9\times 10^{-9}$}
\\ \hline
$\mu$ ($m=2$) & \multicolumn{2}{c|}{$\mu\to e\gamma$ possible} &
\multicolumn{2}{c|}{$5.0\times 10^{-6}$} & 
\multicolumn{2}{c|}{$\tau\to\mu\gamma$ possible}
\\ \hline
$\tau$ ($m=3$) & \multicolumn{2}{c|}{$3.2\times 10^{-9}$} &
\multicolumn{2}{c|}{$\tau\to e\gamma$ possible} & 
\multicolumn{2}{c|}{$\tau\to\mu\gamma$ possible}
\\ \hline
\end{tabular}
\caption{\label{results} 
Branching ratios, scaled by the appropriate RPV couplings and for 
$m_{\tilde\nu_k}=100$ GeV or $m_{\tilde\nu^\star_k}=100$ GeV, 
for each loop particle. Results for quarks with both constituent
and current (in parentheses) masses are given. Also indicated in this 
table are the relevant index combinations for the RPV coupling products 
when either a sneutrino or a charge-conjugate sneutrino is exchanged.}
\end{table*}
\begin{table*}[htb]
\begin{tabular}{|c|c|c|c|} \hline
RPV coupling& \multicolumn{2}{c|}{Current limit} & Expected sensitivity from 
$BR(\mu\rightarrow e\gamma\gamma) < 10^{-14}$ \\ \cline{2-3}
& From \cite{FKKV} & From \cite{Drei,Dr-99} &\\ \hline
$\lambda_{122} \lambda_{233}$ & none & $3\times 10^{-3}$
\footnotemark[3]$^{,}$\footnotemark[4] & $1.8\times
10^{-3}$ \\ \hline 
$\lambda_{121} \lambda_{133}$ & none & $2\times 10^{-4}$
\footnotemark[3]$^{,}$\footnotemark[5] & $1.8\times
10^{-3}$ \\ \hline 
$\lambda_{122} \lambda'_{211}$ & $4.1\times 10^{-9}$ \footnotemark[1] & 
$4\times 10^{-8}$ \footnotemark[6]
& $8.8\times
10^{-4}$ ($4.9\times 10^{-5}$) \\ \hline 
$\lambda_{132} \lambda'_{311}$ & $4.1\times 10^{-9}$ \footnotemark[1] & 
$4\times 10^{-8}$ \footnotemark[6] & $8.8\times
10^{-4}$ ($4.9\times 10^{-5}$) \\ \hline 
$\lambda_{121} \lambda'_{111}$ & $4.1\times 10^{-9}$ \footnotemark[1] &
$4\times 10^{-8}$ \footnotemark[6]
& $8.8\times
10^{-4}$ ($4.9\times 10^{-5}$) \\ \hline 
$\lambda_{231} \lambda'_{311}$ & $4.1\times 10^{-9}$ \footnotemark[1] & 
$4\times 10^{-8}$ \footnotemark[6] & $8.8\times
10^{-4}$ ($4.9\times 10^{-5}$) \\ \hline 
$\lambda_{122} \lambda'_{222}$ & $7.7\times 10^{-9}$ \footnotemark[1]$^{,}
$\footnotemark[2] & $9\times 10^{-3}$ \footnotemark[3]$^{,}$\footnotemark[7]
& $1.4\times
10^{-3}$ ($3.5\times 10^{-4}$) \\ \hline 
$\lambda_{132} \lambda'_{322}$ & $7.7\times 10^{-9}$ \footnotemark[1]$^{,}
$\footnotemark[2] & $0.012$ \footnotemark[4]$^{,}$\footnotemark[8] & $1.4\times
10^{-3}$ ($3.5\times 10^{-4}$) \\ \hline 
$\lambda_{121} \lambda'_{122}$ & $7.7\times 10^{-9}$ \footnotemark[1]$^{,}
$\footnotemark[2] & $2.1\times
10^{-3}$ \footnotemark[3] & $1.4\times
10^{-3}$ ($3.5\times 10^{-4}$) \\ \hline 
$\lambda_{231} \lambda'_{322}$ & $7.7\times 10^{-9}$ \footnotemark[1]$^{,}
$\footnotemark[2] & $0.012$ \footnotemark[4]$^{,}$\footnotemark[8] & $1.4\times
10^{-3}$ ($3.5\times 10^{-4}$) \\ \hline 
$\lambda_{122} \lambda'_{233}$ & none & $7.4\times 10^{-3}$ \footnotemark[3]$^{,}
$\footnotemark[5] & $0.014$
($0.013$) \\ \hline 
$\lambda_{132} \lambda'_{333}$ & none & $0.016$ \footnotemark[4]$^{,}
$\footnotemark[9] & $0.014$ ($0.013$) \\ \hline 
$\lambda_{121} \lambda'_{133}$ & none & $6.9\times 10^{-5}$ \footnotemark[3]$^{,}
$\footnotemark[5] & $0.014$ ($0.013$)\\ \hline 
$\lambda_{231} \lambda'_{333}$ & none & $0.016$ \footnotemark[4]$^{,}
$\footnotemark[9] & $0.014$ ($0.013$) \\ \hline 
\end{tabular}
\caption{\label{lambdas} Current limits and expected sensitivity  
to RPV coupling products from future experiments. Limits are given 
for sparticle masses of $100$ GeV. The limits shown on the 
fourth column scale as $m_{\tilde\nu}/100~[{\rm GeV}]$, see text.}
\footnotetext[1] {From $\mu \to e$ conversion in nuclei.}
\footnotetext[2] {This limit depends strongly on the strange-quark content 
                  in the nucleon and is, therefore, model-dependent. In particular, 
                  if the strange quark content in the nucleon is consistent with zero 
                  (see e.g. \cite{KP-89}), then this limit does not apply.}
\footnotetext[3] {From charged current universality.}
\footnotetext[4] {From $\Gamma(\tau\to
e\nu\bar{\nu})/\Gamma(\tau\to\mu\nu\bar{\nu})$.}
\footnotetext[5] {From neutrino masses.}
\footnotetext[6] {From $\mu \to e$ conversion in nuclei \cite{HMRS}.}
\footnotetext[7] {From $BR(D^+\to\bar{K}^{0*}\mu^+\nu_{\mu})/
                        BR(D^+\to\bar{K}^{0*}e^+\nu_{\mu})$.}
\footnotetext[8] {From $D^0 - \bar{D}^0$ mixing. Model-dependent.}
\footnotetext[9] {From $\Gamma_{had}(Z^0)/\Gamma_{lep}(Z^0)$.}
\end{table*}

A few remarks are in order to conclude the discussion of the branching ratios:

\begin{itemize}
\item There is also a second possibility for $L \to l \gamma \gamma$ of 
      the $\lambda$-{\it irreducible} type to proceed: through the soft 
      SUSY-breaking RPV terms \cite{ADL} 
      $C_{ijk} \tilde{L}_i \tilde{L}_j
       \tilde{E}_k^c$ or $C'_{ijk} \tilde{L}_i\tilde{Q}_j \tilde{D}_k^c$ 
      and slepton or squark in the loop. In this case the sneutrino couples 
      to a pair of sleptons or down squarks of different helicity index, i.e. 
      $\tilde\nu \tilde{f}_L \tilde{f}_R$, where $\tilde{f}$ is a slepton or 
      down squark, while the photons couple to $\tilde{f}_L \tilde{f}_L$ or 
      to $\tilde{f}_R \tilde{f}_R$. Assuming for simplicity that 
      the $\tilde{f}_L \tilde{f}_R$ mixing is of ${\cal O}(1)$ and 
      setting $C_{ijk}\equiv M_{SUSY} \tilde{C}_{ijk}$,
      $C'_{ijk}\equiv M_{SUSY} \tilde{C'}_{ijk}$, the width for 
      $L \to l \gamma \gamma$ is 
      (for $m_{\tilde{f}}^2/m_{L}^2 \gg 1$ and 
           $m_{l}^2/m_{L}^2 \ll 1$):

\begin{widetext}
\bea
\Gamma(L \to l \gamma \gamma) \simeq \tilde{A}_{RPV}^{ij\tilde{f} S}
\left(\frac{\alpha}{\pi}\right)^2 \frac{1}{256\pi^3}
\frac{Q_{\tilde{f}}^4}{8640} 
\frac{m_{L}^7 M_{SUSY}^2}{m_{\tilde{f}}^4 m_{S}^4}~,
\eea
\end{widetext}

where $\tilde{f}=\tilde{\ell}$ with $Q_{\tilde{f}}=-1$ 
   or $\tilde{f}=\tilde{d}$ with $Q_{\tilde{f}}=-1/3$ 
if the sparticle in the loop is a slepton or down squark, respectively, 
and $S=\tilde\nu$ or $\tilde\nu^\star$. Also,

\begin{widetext}
\begin{eqnarray}
\tilde{A}_{RPV}^{ij \tilde{\ell}_m \tilde\nu_k}= 
\mid \tilde{C}_{kmm}\lambda_{kij} \mid^2 ~,~
\tilde{A}_{RPV}^{ij \tilde{\ell}_m \tilde\nu_k^\star}= 
\mid\tilde{C}_{kmm}\lambda_{kji}\mid^2
~,~\nonumber\\
\tilde{A}_{RPV}^{ij \tilde{d}_m \tilde\nu_k}= 
9 \mid \tilde{C'}_{kmm}\lambda_{kij} \mid^2 ~,~
\tilde{A}_{RPV}^{ij \tilde{d}_m \tilde\nu_k^\star}= 
9 \mid \tilde{C'}_{kmm}\lambda_{kji} \mid^2 ~.
\end{eqnarray}
\end{widetext}

Here $m,~k=1,2,3$ denote the family of the slepton ($\tilde{\ell}$) or down 
squark ($\tilde{d}$) and sneutrino or antisneutrino, respectively.

Plugging in numbers 
(e.g., setting $M_{SUSY}=m_{\tilde{\ell}_m}=m_{\tilde\nu_k}=100$ GeV),  
$BR(\mu \to e \gamma \gamma) \sim \mid \tilde{C}_{kmm}\lambda_{k12} \mid^2
 3.7 \times 10^{-14}$ and 
$BR(\tau \to e \gamma \gamma) \sim \mid \tilde{C}_{kmm}\lambda_{k13} \mid^2
 1.9\times 10^{-12}$.
Both values are much smaller than the ones obtained with fermions in the 
loop.

\item If an
      RPV bilinear term exists,
      i.e. $B \tilde{L} H_u$ \cite{ADL}, then $L\to l\gamma\gamma$ can also 
      proceed via the $\lambda$-{\it irreducible} topology with one trilinear 
      ($\lambda$) and one bilinear ($B$) RPV insertion. Therefore, the decay 
      would be $\propto \lambda \times B$. In this case the off-shell sneutrino 
      emitted from the decaying lepton mixes with the down-type Higgs boson, 
      which then goes into the two photons with the well known (see e.g.
      \cite{HHG}) one-loop SUSY amplitude for $H \to \gamma \gamma$ 
      (with the replacement $m_H^2\to2k\cdot k'$). This possibility will be 
      investigated in detail elsewhere \cite{AGt}.
\item Another set of related processes that may be generated through the
      same $\lambda$-{\it irreducible} diagrams are $\tau\rightarrow(e,\mu)gg$.
      Work on these processes is in progress \cite{AGt}.
\item The $\tau$-decays $\tau \to (\mu, e) \gamma \gamma$ of the 
      $\lambda$-{\it irreducible} type that are considered in this paper, 
      with the lepton loop, have tree-level ``analogs'' that probe identical 
      combinations of lambdas: 
      $\tau\to\mu ee$ and $\tau\to\mu\mu e$ \cite{CR-96}.
      However, the contribution from lepton loops is not necessarily the largest 
      one: it was found above that the dominant contribution stems from the 
      $s$-quark loop, for which the branching ratio can amount to up to 
      ${\cal O}(10^{-10})$ (for $\tau\to\mu\gamma\gamma$), taking into account 
      the current constraints \cite{SK-02}, which are obtained from 
      processes of the type $\tau\to l+{\rm hadrons}$.
\item As it can be seen from the Table, $\mu-e$ conversion in nuclei is a 
      strong rival to our processes. However, couplings that involve $b$ quark
      and $\tau$ lepton (and perhaps $s$ quark) are beyond the reach of $\mu-e$
      conversion experiments. Although our bounds on such processes are weak, 
      they will strengthen if PRISM experiment will come into action.  
\end{itemize}

In this paper a new class of lepton flavor-violating-processes, 
$L\to l\gamma\gamma$ has been presented and discussed that might appear even if 
the corresponding decay into one photon only is suppressed or not existing at all.
The underlying mechanism for such decays is the decay of the heavy lepton into
the lighter one and a virtual scalar particle that goes into the two photons
via a triangle diagram. Hence, such processes might probe the flavor
structure of leptons coupling to these scalar particles that are a common
feature of many models for new physics. Branching ratios for such decays have 
been calculated within the framework of R-parity-violating supersymmetry with
sneutrinos as the scalar particles. Imposing RPV bounds emerging from other 
processes, the resulting branching ratios are well below current observation 
thresholds. However, for a planned new round of experiments, especially for 
$\mu$ decays, these processes - if they exist - might shed light on potential
new physics or in turn might help to set more stringent exclusion bounds.

\begin{acknowledgments}
A.G. acknowledges financial support by DAAD and thanks G. Soff and
the members of his group at the Dresden Technical University for 
providing an atmosphere conducive to research  during his visit in Dresden.
S.B.S. would like to thank S. Roy for helpful discussions. This work
is supported by the Israel Science Foundation and by the
VPR Fund for the Promotion of Research at the Technion.
\end{acknowledgments}

\end{document}